\documentclass[a4paper,aps,pre,superscriptaddress,floatfix,nofootinbib,twocolumn,bibtex]{revtex4}

\usepackage{graphicx}
\usepackage{multirow}
\usepackage[english]{babel}

\begin{document}

\title{Universality, limits and predictability of gold-medal performances at the Olympic Games}

\author{Filippo Radicchi}
\affiliation{Departament d'Enginyeria Quimica, Universitat Rovira i Virgili, Av. Paisos Catalans 26, 43007 Tarragona, Catalunya, Spain}


\begin{abstract}
\noindent Inspired
by the Games held in ancient Greece,
modern Olympics represent
the 
world's largest pageant of athletic skill and 
competitive spirit. 
Performances of athletes at the Olympic Games mirror, 
since $1896$, human potentialities
in sports, and thus provide 
an optimal source of information for studying
the evolution of sport achievements 
and predicting the
limits that athletes can reach. Unfortunately, the models introduced so far for
the description of athlete performances at the Olympics are 
either sophisticated 
or unrealistic, 
and more importantly, do not
provide a unified theory for sport performances.
Here, we address this issue by showing that 
relative performance improvements of medal winners at the Olympics
are normally distributed, implying that
the evolution of performance values can be 
 described in good approximation
as an exponential approach to 
an {\it a priori} unknown limiting performance
value. This law holds for 
all specialties in athletics -- including
running, jumping and throwing -- and swimming.
We present a self-consistent method, based
on normality hypothesis testing, able to
predict limiting performance values
in all specialties. We further quantify
the most likely years in which 
athletes will breach challenging performance walls in
running, jumping, throwing and swimming
events, as well as
the probability
that new world records will be
established at the next edition of the Olympic Games.
\end{abstract}

\maketitle

\section*{Introduction}
\noindent  Modern 
Olympics are inspired by the ancient version of the Games, but
based on a wider idea of globality. While
ancient Games were opened only to Greek
speaking athletes~\cite{Swaddling2000}, modern Olympics 
were, since their beginning, 
considered a world event involving people from
every part of the globe~\cite{Guttmann2002}. 
The same symbol of the Olympics, composed of five interlocking rings standing for
the five continents, was  designed
by the {\it Baron Pierre de Coubertin}, the founder of the modern Olympic Games,
with the aim of reinforcing the idea that the Games 
are an international event and welcome all countries 
of the world~\cite{knight92}.
Since Athens $1896$, 
$26$ editions of the event has been organized in different
locations around the world, and, from 
the $241$ participants representing 
$14$ nations of the first edition, 
the Games have grown to about $10,500$ competitors 
from $204$ countries at the latest
edition of the summer Games of Beijing $2008$. The Olympics are
one the most important events worldwide not only
for sports, but also for politics and
society. Many important facts of the last century
history, such as the Nazism~\cite{Mandell1971}, the Israeli-Palestinian 
conflict~\cite{Reeve2000}, and
the cold war~\cite{Guttmann1988}, have influenced
the regular organization of the Games.
Also, the Olympics generally play a fundamental 
and positive role for the economic and urban development of the
city that hosts the event~\cite{Waitt2003, H2004}.
\\
\noindent  Performance data 
of athletes at the Olympics are available for each modern edition
of the Games organized so far, and represent
an optimal proxy for the study of
human limits in sport performances
for three main reasons:
(i) Data cover more than a century of sport performances
since the first edition of the Olympics dates back to $1896$;
(ii) Olympic data provide a detailed record of sports performances at 
regular $4$-year intervals;
(iii) The performances of Olympic medalists truly reflect the
best achievements that could be obtained in a given
historic moment because, in the
vast majority of sport disciplines,
the Games have always represented the most important event
during the career of an athlete, and consequently all the greatest 
athletes have always taken part in the Olympics. 
\\
\noindent Latest years have witnessed the appearance of a large
number of statistical studies of data coming from professional sports.
Examples include basketball~\cite{Ben-Naim2005,Yaari2011}, baseball~\cite{Sire2008,Petersen2008,Saavedra2009,Petersen2010a,Petersen2011}, soccer~\cite{Duch2010}, tennis~\cite{Radicchi2011}, etc.
Also Olympic performance data
have been the subject of many
analyses~\cite{Tibshirani1997,Grubb1998,Sparling1998,Katz1999,Savaglio2000,Holden2004,Atkinson2004,Nevill2005,Nevill2007,Denny2008,Sabhapandit2008}. Some of them focused on models aimed at
the description of performance progression along time, including
linear models~\cite{Atkinson2004} that can even lead
to unrealistic results~\cite{Sharp2004,Rice2004},
S-shaped curves~\cite{Nevill2005} and logistic functions~\cite{Denny2008}.
Others studied statistical properties of performance 
patterns, such as the power-law relation
between time (or speed) and length of running 
events~\cite{Grubb1998,Katz1999,Savaglio2000}.
In addition, performance data of athletes at the Olympics have been
used to tune the parameters of complicated models aimed at the determination of
physiological limits in sport performances~\cite{Sjodin1985,Peronnet1989,DiPrampero2003}.
For example, according to a mathematical model for
human running performance that accounts for
various energetic factors, such as
capacity of anaerobic metabolism, maximal aerobic power
and reduction in peak aerobic power, Perronet and 
Thibault predicted the limiting times that athletes can reach
in  various running events in athletics~\cite{Peronnet1989}.
\\
\noindent  In spite of the numerous efforts however, 
we still miss a general description
for the performances of athletes. We still
miss a universal way to predict limiting performance values
and calculate the probability of future achievements in sport.
In this paper, we address all these issues by generating
a simple and coherent picture for the description of the performances
obtained by Olympic medal winners in 
all specialties of athletics and swimming.
We analyze historic performance
data and  provide empirical evidence about the discovery of a 
novel statistical law governing performances of medal winners 
at the Olympic Games. 
With a self-consistent approach we simultaneously 
(i) show that performance improvements 
obey a universal law, 
(ii) estimate limiting performance values,
(iii) predict future achievements at the Olympics.

\section*{Results}
\noindent  While former statistical studies have mainly
analyzed the progression of absolute
performance values along the various
editions of the Games, here we change point of view and focus our
attention on relative improvements in performances
between two consecutive editions of the Olympics.
Let us indicate with $p_y$ the value of the performance
obtained by the gold medalist in a specific specialty
at the edition  of year $y$ of the Olympic Games. Depending on the
specialty, $p_y$ may indicate
time (running and swimming), length (long and triple jumps), height
(high jump and pole vault), or distance (discus and hammer throws, 
shot put). We define the relative
improvement of the gold-medal performance in the Games
of year $y$ with respect to the gold-medal performance
in the previous edition of the Olympics as
\begin{equation}
\xi_y \; \colon = \; \left( \Delta p_{y-4} - \Delta p_{y} \right) / \Delta p_{y-4}  \;\; ,
\label{eq:1}
\end{equation}
where $\Delta p_y = p_y - p_\infty$
represents the gap between the performance
value of the gold medalist in year $y$ and the asymptotic
performance value $p_\infty$.
The asymptotic or limiting performance value $p_\infty$ 
is a unknown parameter representing
the physiological limit that can be achieved in the specialty
by an athlete.
Eq.~\ref{eq:1} defines the relative improvement
towards the asymptotic performance value of the gold
medalist in year $y$ with respect to the performance
of the gold medalist in year $y-4$.
Note that the same definition can be
used for the measurement of the relative improvements
of silver and bronze medalists, and in principle for athletes who have 
reached any arbitrary rank position.
\\
For reasonable values of $p_\infty$, we find that the distribution
of the relative performance improvements
is statistically consistent with a normal distribution. We determine the best
estimate of the asymptotic performance value
$\hat{p}_\infty$ as the value of $p_\infty$
for which the statistical significance ($p$-value)  of 
the normal fit is maximized (see Materials and Methods section). The procedure is 
generally  accurate and allows
us to identify reasonable values of 
$\hat{p}_\infty$ in all specialties considered in this
study. In Fig.~1 for example,
we report the results obtained by analyzing performance data
of male athletes in $400$ meters sprint. The best
estimate of the asymptotic time is $\hat{p}_\infty=41.62$ seconds. 
For this value of $p_\infty$, we find that relative
improvements obey a normal distribution with
average value $\hat{\mu}=0.06$ and standard deviation
$\hat{\sigma}=0.19$. 
Statistical significance, however, can be used not only for the determination
of the best estimate of the asymptotic performance value, but also, in a broader sense,
to define confidence intervals for $\hat{p}_\infty$.
In the case of $400$ meters sprint
of male athletes for example, we find that, at $5\%$ 
significance level, $\hat{p}_\infty$ is in the range $31.03$ to $43.09$ seconds. 
At $50\%$ significance level, the
interval is restricted and $\hat{p}_\infty$ is 
in the range $38.91$ to $42.74$ seconds,
while, at $95\%$ significance level, $\hat{p}_\infty$ is expected
to be between $41.04$ and $42.13$ seconds.
The results shown in Fig.~1 are obtained by analyzing the relative
performance improvements of gold-medal winners. Similar
results are, however, obtained when considering the performances
of silver and bronze medal medalists (Fig.~S1). 
Interestingly, the finiteness of the data
does not affect the reliability of the
best estimate of the limiting
performance value since compatible values of $\hat{p}_\infty$
can be detected by removing results of the latest editions of the Games from the analysis
 (Fig.~S2).
\\

\begin{figure}[htb]
\includegraphics[width=0.45\textwidth]{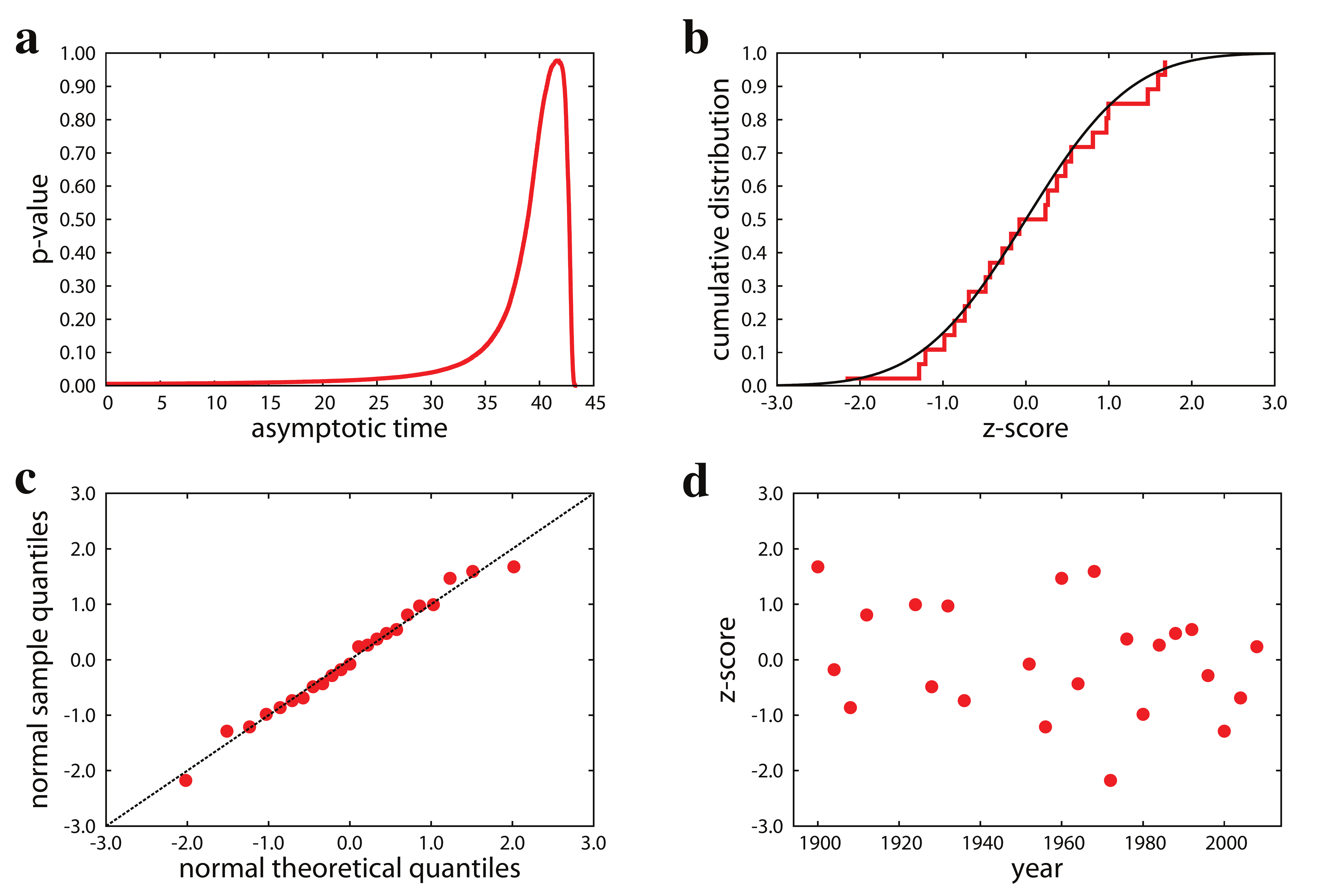}
\caption{{\it Performances of male gold medalists in 400 meters sprint.}
{\bf a.} Best estimate of the
asymptotic performance value. For each value of $p_\infty$ lower than the
actual Olympic record, we evaluate the goodness
of the fit of performance improvements with a
normal distribution. $\hat{p}_\infty$ is determined as
the value of the asymptotic time $p_\infty$ that maximizes the 
statistical significance ($p$-value).
For men $400$ meters sprint, our best estimate is 
$\hat{p}_\infty = 41.62$ seconds,
where we find that
relative performance improvements 
are normally distributed with a confidence
of $98\%$. For this value of $p_\infty$, 
 the best empirical estimates of the
average value and standard deviation are
respectively $\hat{\mu} = 0.06$ and
$\hat{\sigma} = 0.19$. {\bf b.} The cumulative
distribution function of the $z$-scores obtained 
for $p_\infty=\hat{p}_\infty$ (red curve) is compared with the standard normal
cumulative distribution (black curve). 
{\bf c.} Normal sample quantile are plotted against normal 
theoretical quantiles~\cite{Wilk1968}.
The dashed line corresponds to the theoretically
expected behavior in case of a perfect agreement between sample
and theoretical distributions. 
{\bf d.} $z$-scores of relative
performance improvements between consecutive
editions of the Games.}
\end{figure}

\begin{figure}[htb]
\includegraphics[width=0.45\textwidth]{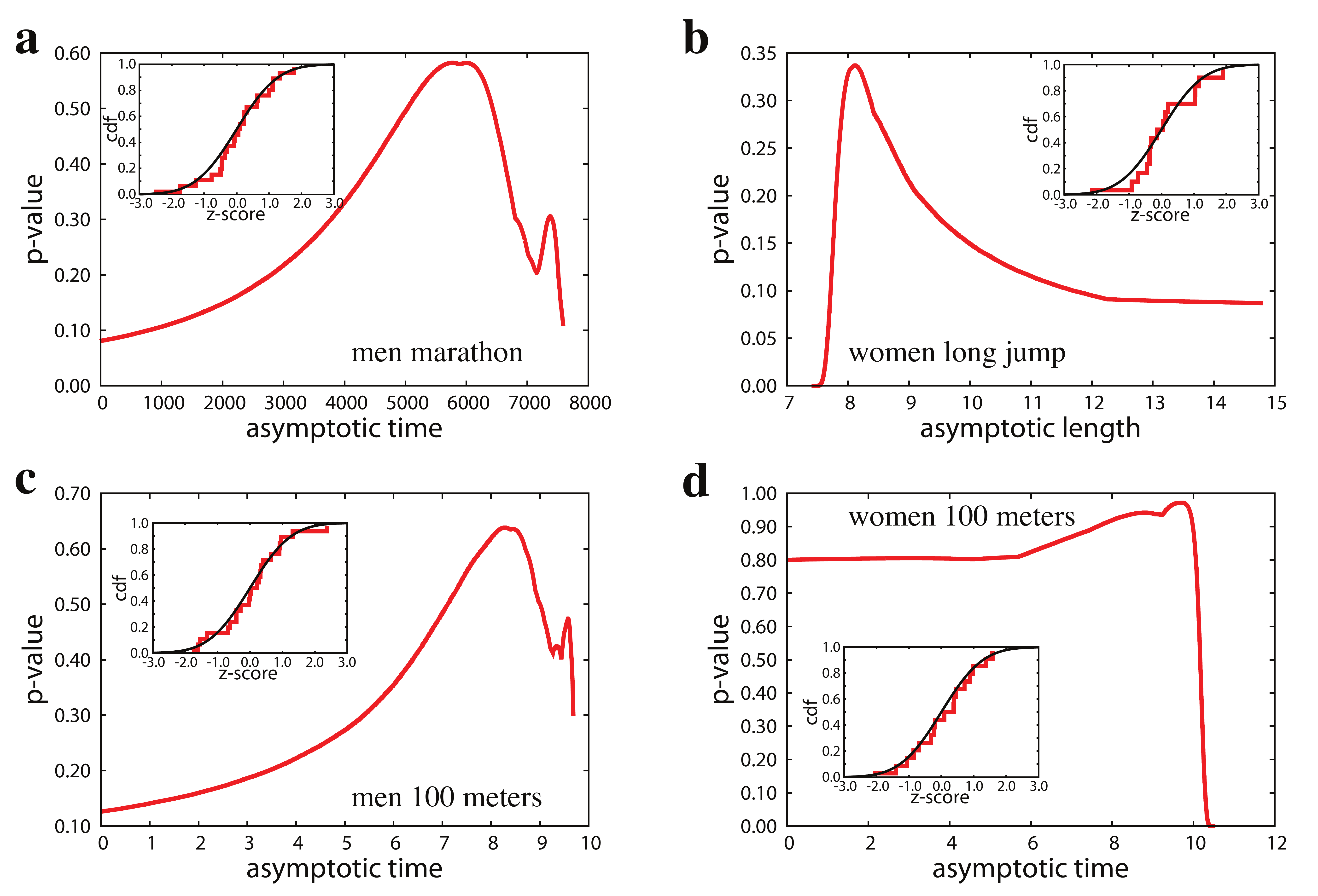}
\caption{{\it Statistical properties of performance 
improvements in athletics.}
In the main panels we show the determination of the best
estimate $\hat{p}_\infty$ of the asymptotic performance value, while 
in the insets we provide a graphical comparison
between the sample cumulative distributions (red line) and the
standard normal cumulative distribution (black line). 
{\bf a} and {\bf b.} We report
the results obtained by the analysis of the performances
of male athletes in marathon 
($\hat{p}_\infty = 5,771.44$ seconds, $p$-value $= 0.58$) 
and female athletes in
long jump ($\hat{p}_\infty = 8.12$ meters, $p$-value $= 0.34$). 
{\bf c} and {\bf d.} We show 
the outcome of our method for performances of
men and women in $100$ meters sprint (respectively, 
$\hat{p}_\infty = 8.28$ seconds and $p$-value $= 0.64$, 
$\hat{p}_\infty = 9.72$ seconds 
and $p$-value $= 0.97$). }
\end{figure}

\noindent The normality of the relative improvements
towards the asymptotic performance value is a 
simple and strong result.
At each new edition
of the Games, gold-medal performances get, on average, closer to
the limiting performance value.
The average positive improvement observed in historic 
performance data can be motivated
by several factors: as time goes on, athletes are becoming more professionals,
better trained, and during the season have more events to participate in;
the pool for the selection of athletes grows with time, and, 
consequently there is a higher level of competition; 
the evolution of technical materials favors better performances. 
On the other hand, there is also a non null 
probability that winning performances
become worse than those obtained in the previous edition of the Games 
(i.e., relative improvement values are negative). 
All these possibilities are described by a
Gaussian distribution that accounts for
various, in principle hardly quantifiable, factors that may influence
athlete performances: meteorological
and geographical conditions,
athletic skills and physical condition of the
participants, etc. 
The accuracy of the
normal fit is not only testified by
its high statistical significance, but also
by graphical comparisons between the sample distribution
 and the theoretical normal distribution
(see Figs.~1b and c). It is also important to note
that the values of the relative improvements 
do not depend on the particular edition
of the Games, and thus their distribution is stationary (Fig.~1d).
The strength of our results, however, is not only in the
significance of the fits, but especially in its generality.
We repeated the same type of analysis for a
total of $55$ different specialties, and 
found that performance improvements are governed by a
universal law.
First of all, the law holds for all running events in athletics. 
This is valid for an heterogeneous set of running
distances ranging from $100$ to $42,195$ meters 
(marathon, Fig.~2 and Supporting Information S1). Second,
our analysis suggests that relative improvements
are normally distributed not only
when considering time performances, but also 
performances regarding length or height (jumps) and 
distance (throws).
In Fig.2b for example, we report the outcome
of our method when applied to
performance data of female gold medalists
in long jump. Other examples can be found in Supporting Information S2.
Finally, the law is valid for performance improvements of athletes in
swimming specialties (Supporting Information S3).
\\
\noindent Given the 
attention received in the recent 
past~\cite{Atkinson2004,Sharp2004,Rice2004}, 
we reserve a special consideration to the comparison 
in performances between female 
and male athletes in $100$ meters sprint. 
In Fig.~2c and~2d, we report
the results obtained through the analysis of
Olympic performances in this specialty. According to our
analysis, the best estimate of the limiting time for males
is $\hat{p}_\infty = 8.28$ seconds, while for females
we identify the best estimate for the asymptotic time
at $\hat{p}_\infty=9.72$ seconds. Our statistical analysis
predicts that women will be always slower
than men and that the gap will saturate at about $14\%$,
consistent with the estimation by Sparling {\it et al}~\cite{Sparling1998} but
in disagreement with what predicted by 
the unrealistic model of Atkinson {\it et al}~\cite{Atkinson2004}.
It should be noted that 
for women the statistical significance is 
less predictive than the one measured for men. While
for men we observe that statistical significance is 
clearly peaked around $\hat{p}_\infty$ and goes rapidly to zero
as $p_\infty$ decreases, the same does not
happen in the case of women. We believe that the statistics are less 
accurate because the analysis
is based on $19$ editions instead of 
$26$ since women started to run the $100$ meters sprint 
only in Amsterdam $1928$,  while men already in Athens $1896$.
In particular, the lack of
sufficient data provides high statistical significance also
for the unrealistic $p_\infty=0$ seconds.
We expect, however, that the future addition of more data point
will suppress this effect. Despite these
problems, our analysis still produces meaningful 
estimates of the upper bound of the asymptotic time:
at $5\%$ significance level, the asymptotic value
is expected to be lower than $10.31$ seconds, while
at $50\%$ significance level,  $\hat{p}_\infty$ should
be lower than $10.17$ seconds.
Also, our best estimates of the limiting
performance values are probably 
not as accurate for this specialty 
(or other short distances) 
because there is not enough reliable performance
data regarding the first editions of the Games (automatic time
was introduced in Mexico City $1968$). 
The removal
of data points for male $100$ meters sprint before Amsterdam $1928$
(and in general of a few data points from the entire time serie)
leads also to the impossibility to
determine the best estimate of the asymptotic time
as a global maximum of statistical significance (see Fig.~S3).
For $100$ meters sprint, we have performed therefore
an additional analysis in which we aggregated
together the results of gold, silver and bronze medalists
and obtained slightly different estimates for the limiting performance
values [$\hat{p}_\infty=8.80$ seconds for men (Fig.~S4)
and $\hat{p}_\infty=9.64$ seconds for women (Fig.~S5-S6)].
\\

\begin{figure}[htb]
\includegraphics[width=0.45\textwidth]{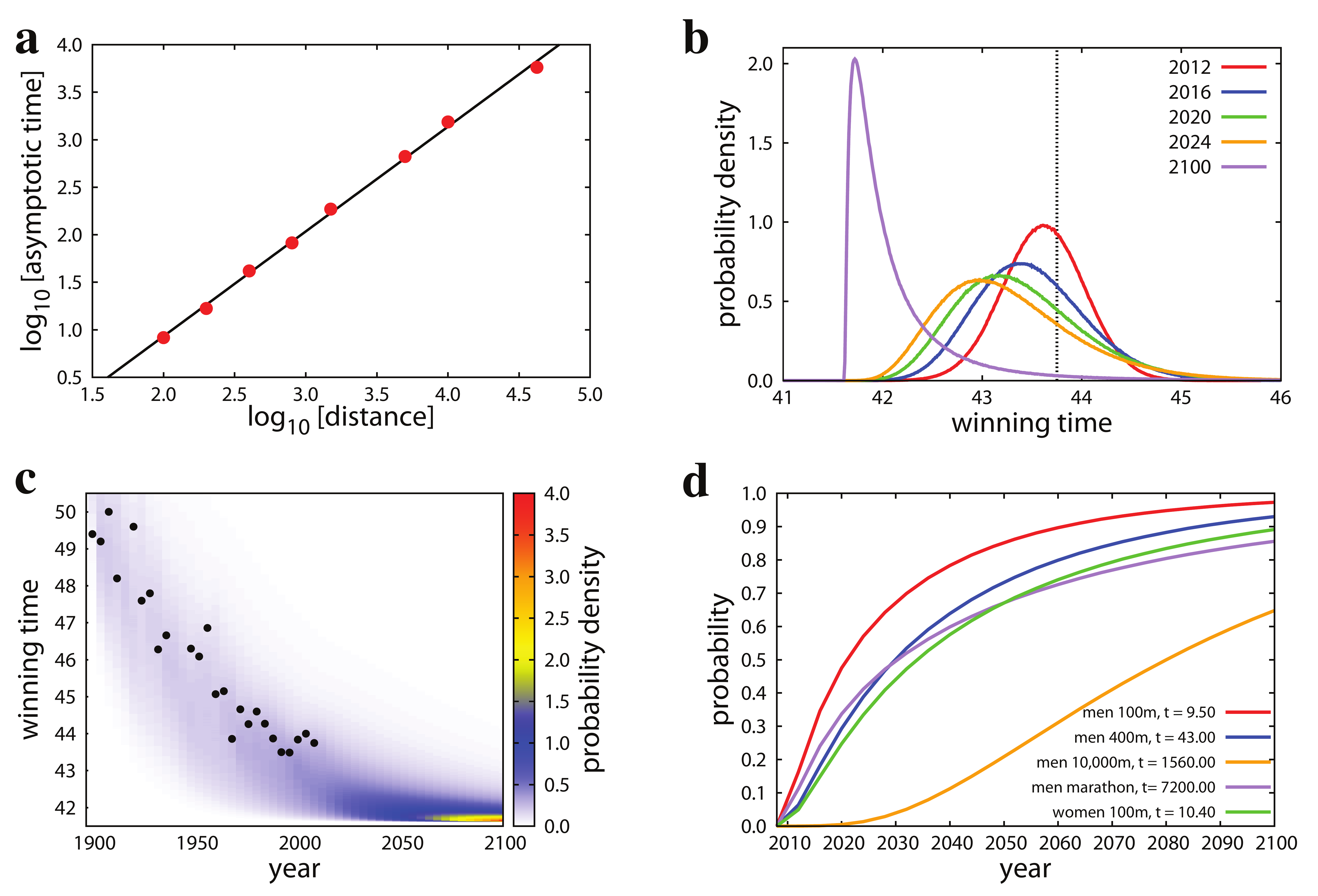}
\caption{{\it Scaling
law between asymptotic time and
running length, and prediction of performances at future 
editions of the Olympic Games.}
{\bf a.} Relation between the best estimates of the limiting performance
value $\hat{p}_\infty$ and the length $\ell$ of the race for 
men running events in athletics (red circles). We excluded
from the analysis relay and
hurdles events. We find that
$\hat{p}_\infty \sim \ell^{\alpha_\infty}$, and
the best estimate of the
power-law exponent is $\hat{\alpha}_\infty=1.10 \pm 0.02$ (black line).
{\bf b.} Probability density functions of the winning time for the
men $400$ meters sprint in 
future editions of the Games. The dashed line represents the winning time
in the latest edition of the Olympics in Beijing $2008$. This value is used
as initial condition for the
prediction of future performances.
{\bf c.} The probability density of the winning time in men $400$ meters
predicted by our model is compared to past performance data (black circles).
The density plot is obtained by convoluting the various
prediction curves derived from real data.
{\bf d.} Probability that athletes will breach challenging walls
in various specialties of athletics as a function of time.}
\end{figure}

\begin{table*}

\begin{center}
\begin{tabular}{llrrrrrrrc}
sport & gender & specialty & $\hat{p}_\infty$ & $\hat{\mu}$ & $\hat{\sigma}$ & $p$-value  & $E$ & $P$ & $\hat{p}_{2012}$\\
\hline \hline
\multirow{10}{*}{Track \& Field} 
& \multirow{7}{*}{Men} 
& 100m & 8.28   & 0.04 & 0.10	& 0.64  & 26  & 0.35 & 9.63 $\pm$ 0.13\\ \cline{3-10}
& & 110m hurdles & 11.76  & 0.05 & 0.12 & 0.48	&  26 & 0.50 & 12.87 $\pm$ 0.14\\ \cline{3-10}
& & 400m & 41.62  & 0.06 & 0.19	& 0.98  &  26 & 0.14 & 43.62  $\pm$ 0.41\\ \cline{3-10}
& &  10,000m & 1,539  & 0.05 & 0.19	& 0.45  &  22 & 0.01  & 1,617 $\pm$ 15 \\ \cline{3-10}
& & marathon & 5,771  & 0.03 & 0.15	& 0.58  & 26 &  0.34 &  7,537  $\pm$ 273 \\ \cline{3-10}
& & pole vault & 6.87  & 0.05 & 0.08	& 0.91  & 26 &0.03 & 6.00 $\pm$ 0.07 \\ \cline{3-10}
& & hammer throw & 103.81 & 0.04 & 0.09	& 0.47  & 25 &0.03 &  82.89 $\pm$ 1.96 \\ \cline{2-10} 
& \multirow{3}{*}{Women}
& 100m & 9.72 & 0.05 & 0.19	& 0.97  & 19 & 0.12 & 10.73  $\pm$ 0.20 \\ \cline{3-10}
& & 400m & 45.14 & 0.02 & 0.15	& 0.77  & 12 & 0.00 & 49.53 $\pm$ 0.67 \\ \cline{3-10}
& & long jump & 8.12 & 0.04 & 0.18	& 0.34  & 16 & 0.01 &  7.08  $\pm$ 0.19\\
\hline \hline
\multirow{8}{*}{Swimming} 
& \multirow{4}{*}{Men} 
& 100m fs & 44.84   & 0.09 & 0.10	& 0.92  & 23&0.36 & 47.00 $\pm$ 0.24 \\ \cline{3-10}
& & 100m bs & 48.98   & 0.09 & 0.11	& 0.93  & 22 &0.24 &  52.22  $\pm$ 0.39 \\ \cline{3-10}
& & 100m brs & 57.38   & 0.16 & 0.16	& 0.93  & 11 & 0.36 & 58.67 $\pm$ 0.24 \\  \cline{3-10}
& & 1,500m fs & 577  & 0.05 & 0.05	& 0.50  & 23 & 0.71 & 866 $\pm$  15 \\ \cline{2-10}
& \multirow{4}{*}{Women} 
& 100m fs & 51.87   & 0.12 & 0.19	& 0.54  & 22 &0.00 &  52.97  $\pm$ 0.24 \\ \cline{3-10}
& & 100m bs & 54.73  & 0.08 & 0.14	& 0.59  & 20 &0.20 & 58.62 $\pm$ 0.59 \\ \cline{3-10}
& & 100m brs & 62.08   & 0.13 & 0.10	& 0.86  & 11 &0.15 & 64.77 $\pm$ 0.31 \\ \cline{3-10}
& & 800m fs & 388  & 0.05 & 0.07	& 0.84  & 11 &0.76 & 489  $\pm$ 7 \\ 
\hline
\hline
\end{tabular}
\end{center}
\caption{{\it  Predictions of gold-medal performances in athletics and swimming.}
We summarize here some of the results
obtained with our analysis. We list several specialties in
athletics and swimming performed by male and female athletes.
For each specialty, we report from left to right: 
the name of the specialty, the
best estimates of the asymptotic performance value $\hat{p}_\infty$,
the best estimate of the mean value $\hat{\mu}$,
the best estimate of the standard deviation $\hat{\sigma}$,
the statistical significance or $p$-value of the 
test of normality, the number $E$ of Olympic Games that
included the specialty, the probability $P$
that the actual world record will be beaten in
London $2012$, and the most likely
performance value $\hat{p}_{2012}$ that gold-medal winners
will obtain
at the next edition of the Olympic Games.
For shortness of notation, in swimming
specialties we abbreviate ``freestyle'' with
``fs'', ``backstroke'' with ``bs'',
and ``breaststroke'' with ``brs''.
The values of $\hat{p}_\infty$ and $\hat{p}_{2012}$ are reported in seconds
for running and swimming races, and in meters for
jumping and throwing events.}
\end{table*}

\noindent In general, our approach produces good results
for specialties with a sufficiently long
tradition in the Games. This is basically the case of all male specialties
in athletics. Data about female performances
typically provide less accurate results, but still, in the majority
of the cases, the predictions of the asymptotic performance values
are reasonable.
We summarize in Table~1 the results obtained 
for some specialties, while
we refer to the Supplementary Information
for a systematic
analysis of all of them.
It should be noted that there are also 
a few cases in which things do not work perfectly.
In women 800 meters, for example, statistical significance
does not exhibit any peak value (Supporting Information S1).
There are also a few specialties
in which the best estimate
of the limiting performance value does not correspond
to the global maximum of  statistical significance 
(Supporting Information S1).
In these cases, statistical significance
is a non monotonic function of the $p_\infty$ and
more maxima are present. Still the peak value
 that appears more plausible can be used as an estimate
 of $\hat{p}_\infty$.
Finally, there are three specialties in athletics in which a clear peak
in statistical significance
is visible only by excluding performance data
of Sidney $2000$, but this exclusion is fully justified
by the fact that the top
athletes of the moment did not take part in 
the competition (Supporting Information S1).
For example, about the men 200 meters sprint
of Sidney $2000$, the web site {\tt sports-reference.com} 
reports: ``This race was expected to be between the 
Americans {\it Maurice Greene} and {\it Michael Johnson}. 
Greene was the best in the world at $100$ meters and 
Johnson at $400$ meters, and their race in the middle 
distance was highly anticipated. But neither qualified 
for the team at the Olympic Trials, succumbing to minor 
injuries, although they both made the team in their better events."
\\
The good accuracy of our best estimates
of the limiting performance values is supported also
by the power-law relation between these quantities
and the length of the
running events in athletics (see Fig.~3a). 
As already observed by Katz and Katz,
world record times ($p_{wr}$) and running distances ($\ell$) are related
by the power-law relation $p_{wr} \sim \ell^{\alpha}$~\cite{Katz1999}.
Katz and Katz studied the relation between world
record performances and running distances in various epochs, and found
that the power-law exponent value $\alpha$ is 
always slightly larger than $1.1$
but decreases for more recent epochs. For example, they measured 
$\alpha \simeq 1.14$ in $1925$, and $\alpha \simeq 1.12$ in $1995$.
On the basis of our measurements, 
we claim that the asymptotic value
of the exponent will be exactly $\alpha_\infty = 1.1$,  when
limiting performance values,
and thus definitive world records, will be reached in all specialties
of athletics.
\\
\noindent A final application of
our findings is the prediction of
future performances at the Olympics. 
The performance value of the gold medalist
in London 2012, for example, can be estimated as 
$p_{2012} = \left( p_{2008} - \hat{p}_\infty\right) \left( 1-\xi \right) + \hat{p}_\infty$ , where 
$\xi$ is a random variate extracted from the normal distribution
$\mathcal{N} \left( \xi; \hat{\mu}, \hat{\sigma} \right)$ with
mean value $\hat{\mu}$ and standard
deviation $\hat{\sigma}$. Similar equations can be written
also to predict performance values of the other editions
after London $2012$. 
For each future edition of the Games,
we can draw a distribution of performance values (see Fig.~3b). 
The distribution
is normal for the edition of $2012$, but diverges from normality
as time grows. In particular, 
while the expected performance value decreases exponentially towards
the asymptotic performance value as time increases, 
the standard deviation initially grows
as we move further in future until predictions become again more accurate
because of the boundary effect of $\hat{p}_\infty$ 
(see Fig.~3c).
\\
By simply looking at the
performances expected at the next edition
of the Games in London $2012$, we can ask what
is the probability
that the winner of the gold-medal
will beat the actual world record of her/his
specialty. In Table~1,
we list these probabilities for some specialties together with
the most likely performance values
that gold-medal winners will obtain. In athletics,
there are not negligible chances (about $30\%$)
that the actual world records of $100$ meters, $110$ meters hurdles
and marathon will be lowered by men. 
In swimming specialties, the expectations are more
promising: there is a good probability (higher than $70\%$)
that the world record of
1,500 meters freestyle will be beaten by male athletes.
\\
Relevant limits are unlikely to be broken
at the next Olympics (Fig.~3d). We will have to wait until $2020$ 
in order to have a $50\%$
chance that a man will run the $100$ meters in less than $9.50$ seconds. 
For other specialties, expectations (probability
higher than $50\%$)
are even less promising: 
men will run the $400$ meters in less than $43.00$ seconds 
and the marathon in less than two hours ($7,200$ seconds) only
after $2030$, women will run the $100$ meters sprint
in less than $10.40$ seconds only after $2040$,
and finally the wall of $26$ minutes ($1,560$ seconds) in
$10,000$ meters will likely be breached by male athletes only after year 
$2080$.

\section*{Discussion}
\noindent  In conclusion, our paper
shows that the performance of
Olympic medal winners in athletics and swimming
obey, independently
of the type of specialty, a simple universal law. 
If performance improvements are calculated
with respect to an asymptotic performance value,
then the relative difference between improvements
obtained in two different editions of the Games
is a random variate following a normal
distribution. This is the common property of a broad class of natural phenomena
that be
described by the theory of biased random walks~\cite{Redner2001}, such as
the locomotory movements of organisms 
responding to an external stimulus~\cite{Alt1980, Hill1997,Codling2008},
the activity of spiking neurons~\cite{gerstein1964}, the trends of
daily temperatures~\cite{Wergen2010}, stock
prices~\cite{Wergen2011}, capital markets~\cite{Peters1989}, etc.
\\
The normality of the relative improvements cannot be explained in trivial
terms, especially in this case where the statistics is performed
on extremal properties of the system. Remember
in fact that the performance values
analyzed here are those obtained by the best athletes
of a given edition of the Olympics (i.e., potentially
the best performers on the earth), and thus it is natural to expect
that absolute performance values obey 
statistical laws of extremes~\cite{Beirlant2004}.
More importantly, since the distribution is normal, it makes sense to refer
to average trajectories of top performance values along 
editions of the Games. Our findings in fact allow to say
that, on average, the absolute performance
value of top athletes at the Olympics gets closer
to the limiting performance value in an exponential fashion,
with a rate of about $5\%$ in athletics and
$10\%$ in swimming. More in detail, the
average trajectory of the performance value
can be described by the equation
\begin{equation}
\left\langle p_y \right\rangle = p_{y_0} \; e^{-\hat{\mu} \left(y-y_0\right)} + \hat{p}_\infty\;\;,
\end{equation}
where $y_0$ is an arbitrary
initial edition year of the Olympics and $p_{y_0}$ is the
performance value measured in year $y_0$. Eq.2 can be derived
directly from Eq.1 and the fact that relative improvements are normally distributed but
only under the assumptions that the edition year of the Olympics is 
considered as a continuous variable
and that $\left\langle \frac{d \Delta p_y / dy} {\Delta p_y} \right\rangle  =  \frac{d \left\langle \Delta p_y \right\rangle /dy}{\left\langle \Delta p_y \right\rangle}$. Note that this 
observation is important for stressing
the difference between our fitting procedure and a more straightforward analysis based on the
exponential fit of absolute performance values, as the one used to find
that the progression
of world record performances
follows a piecewise exponential decaying pattern~\cite{Berthelot2008,Desgorces2008,Guillaume2009}.
Note also that the analysis of 
the only Olympic performances differs from the
one of world record performances for the following reasons: (i) The relative
change between two world records, if defined
in a similar manner as Eq.1, can be only a
positive quantity; 
(ii) The time difference between two world record performances
is not a constant, but a random variate by itself. Because the number
of events in which new world records can be established
is higher today than it was one century ago 
(and they had been growing in the course of the years), in
any analysis of the progression of
world record performances time should be rescaled 
to account for that~\cite{Berthelot2008}.
\\
The asymptotic performance value $p_\infty$
is an {\it a priori} unknown variable whose
value can be self-consistently determined by
maximizing the statistical significance
of the normality fit. 
It is particularly important to stress that our 
simple methodology
provides good estimates of performance limits that are
in general consistent with those obtained
through complicated physiological 
models~\cite{Sjodin1985,Peronnet1989,DiPrampero2003}.
For example,
Perronet and Thibault predicted that the limiting time for men
in marathon is $1$ hour, $48$ minutes and $26$ seconds~\cite{Peronnet1989}.
With our minimalistic model, we are able
to predict that this limiting time is
between $1$ hour, $36$ minutes and $11$ seconds
and $1$ hour, $41$ minutes and $40$ seconds (for men marathon
the peak of statistical significance is wide, see Fig.~2a).
At the same time, it is also important to stress
that our minimalistic analysis can also lead 
to little inconsistencies. For example, the best estimates
of $p_\infty$ obtained here state
that, asymptotically, the average pace in marathon would 
be higher than the one
in $10,000$ meters. This means that according 
to our estimates, the first $10,000$
meters in marathon would be run in 
less than $23$ minutes, 
while the entire race of $10,000$ meters 
would be run asymptotically in more than $25$ minutes.
This inconsistency can be partially explained by the fact that
the statistics for $10,000$ meters is 
less reliable because based only on
$22$ events, while the one for marathon on the results of $26$ 
editions of the Games. In general, it is very important
to remark that, at the moment, we are
able to provide only good estimates of the asymptotic
performance values because such estimates are based 
on a relatively small set of empirical data (at best
$26$ editions of the Olympics), and therefore must
be taken with a grain of salt.
We expect in fact that, while the 
normal law governing performance improvements will 
likely continue to hold, 
the accuracy in the
estimation of the asymptotic performance
values will improve with the addition of more data points in the future,
starting already from the next edition of the Games in London $2012$.

\section*{Materials and Methods}

\subsection*{Data set}
\noindent Medal lists and
results of all editions of the
Olympic Games have been collected from  
the web sites {\tt www.sports-reference.com}
and {\tt www.databaseolympics.com}. Whenever possible,
we considered automatic measures of time instead of
manual ones.
We included in our study all results obtained in the editions
of the modern Olympic Games since Athens 1896, but 
we excluded from the analysis data about the so-called
``Intercalated'' edition of the Games held
in Athens in $1906$. We focused on
sports classified as ``Track \& Field'' and ``Swimming'',
and particularly
on specialties of these sports that have
been performed at least in the latest 
ten editions of the Olympic Games.
We compared only performances between subsequent editions
of the games held at four years of difference. We excluded
therefore comparisons between either the consecutive editions of
Stockholm $1912$ and Antwerp $1920$ (separated
by World War I), and those of Berlin $1936$ and London $1948$
(separated by World War II).
\\
For consistency, we considered only specialties
whose rules or techniques have not changed during time.
For example, we excluded javelin throw because of the javelin
redesign in $1986$. We also excluded performances in high jump
before Mexico City $1968$ when athletes started for the first
time to adopt the modern jump style called ``Fosbury flop''.
\\
Data  are made available for download at
{\tt filrad.homelinux.org/resources}.

\subsection*{Normality test}
\noindent The results reported in the paper are based
on the normality test introduced by Anderson and Darling~\cite{Anderson1952}.
Given a value of $p_\infty$, we compute the best estimates of
the mean $\hat{\mu}$ and the standard deviation $\hat{\sigma}$ as
$\hat{\mu}  = 1/R \; \sum_y \xi_y$ and  $\hat{\sigma} = \sqrt{1/\left(R-1\right) \; \sum_y \left(\xi_y - \hat{\mu} \right)^2}$, respectively.
The relative improvement $\xi_y$ is defined in Eq.1. $R$ indicates the number of 
results between consecutive editions of the Olympic Games 
that are included in the analysis.
We then compute the $z$-scores as $z_y = \left(\xi_y - \hat{\mu} \right) / \hat{\sigma}$ 
and rearrange them in 
ascending order such that $z_1\leq z_2 \leq \ldots \leq z_R$.
The Anderson-Darling distance is computed with the
formula
$A^2 = -R - 1/R \, \sum_{i=1}^R \left[\left(2i-1\right) \log{\Phi\left(z_i\right)} + \left(2(R-i)+1\right) \log{\left(1-\Phi\left(z_i\right)\right)} \right]$, 
where $\Phi\left(z_i\right)$ is the standard normal cumulative
distribution function.
We further use the modified statistics
$A^{*2} = A^2 \, \left(1 + 4/ R - 25/R^2\right)$, suitable
in the case in which both the mean and standard deviation are 
estimated from the data as
suggested by Stephens~\cite{Stephens1974}. 
\\
We evaluate the goodness of the fit
by generating $10^5$ random number sequences of length $R$ extracted
from the standard normal distribution. 
The statistical significance of the normality
test ($p$-value) is calculated as the
number of artificial sequences  whose $A^{*2}$ is larger than 
the one measured for real data divided
by the total number of generated sequences. Note that
there is a trivial monotonic relation between the $p$-value
and the Anderson-Darling distance  $A^{*2}$, and therefore
the maximum of the $p$-value corresponds to the
minimum of $A^{*2}$.
\\
We used the normality test by Anderson and Darling
because this test is considered one of the best empirical distribution 
function statistics for detecting most departures from 
normality, and can be used
for testing the normality of very small sample sizes~\cite{Stephens1974}.
We verified, however, the robustness 
of our results by using
other standard normality tests, including 
those based on the criteria of Kolmogorov-Smirnov, Cram\'er-von Mises
and Shapiro-Wilk~\cite{Shapiro1965, D'Agostino1986}.
We also verified the consistency of our results with
normality tests based on the moments of the distributions (see Fig.~S6).
\\
Furthermore, we tested the accuracy of our fitting
method by implementing a bootstrap procedure~\cite{Efron1993}, and found
that our fitting method
is able to well recover the correct  parameter values in artificial sequences
generated according to our model (see Fig.~S7).

\section*{Acknowledgments}
\noindent We thank C. Castellano, P.S. Dodds, E. Ferrara and A. Hockenberry
for comments and suggestions on the manuscript.

\end{document}